# Computer Modeling of Irregularly Spaced Signals. Statistical Properties of the Wavelet Approximation Using a Compact Weight Function

Ivan L. Andronov, Violetta P. Kulynska

Department of Mathematics, Physics and Astronomy
Odessa National Maritime University, Odessa, Ukraine

*Abstract*

The algorithm of modified wavelet analysis is discussed. It is based on the weighted least squares approximation. Contrary to the Gaussian as a weight function, we propose to use a compact weight function. The accuracy estimates using the statistically correct expressions for the least squares approximations with an additional weight function are compared with that obtained using the bootstrap method.

*Introduction*

Wavelet transform is a method, which is widely used in science, and in astronomy as well. Typically, in a time series analysis, the argument $t_k$ of a signal $x_k$ ($k = 1..n$) is interpreted as "time". However, it may have another sense in the dependence $x(t)$.

The classical method is based on the analysis of infinite continuous function, so the infinite number of observations (e.g. [1]). In reality, only a limited number of observations is available. In the simplest case, the times are regularly distributed $t_k = t_j + (k - j) \cdot \delta$, $(k, j = 1..n)$, where δ is a time step (time resolution), so the coefficients may be computed using the Fast Fourier Transform (FFT) [2]. Such an algorithm is implemented into some software packages.

Generally, the discrete data are distributed very irregularly, e.g. in photographic, visual or CCD photometric surveys from ground based or space observatories. This challenges oversimplified formulae obtained for infinite data. The corresponding reviews on subsequently increasing number of methods are presented in [3-5].

*Basic formulae*

The mathematical model for the wavelet analysis of discrete signals with generally irregular times is described by Andronov [6,7]. Contrary to the direct replacement of the integrals (valid for a signal infinite in length) by sums [8], the least squares (LS) version is equivalent to that using orthogonal functions [6,7,9]. The signal to noise ratio SNR may be increased by a factor of many times, if using the least squares method instead of oversimplified formulae [6,7].

The approximation of the signal is
$$x_C(t) = C_1 - C_2 \cos(C_4 \cdot (t - C_3)), \tag{1}$$

where $C_1$ — the mean value of the approximation during the period (generally, not coinciding with a sample mean), $C_2$ — is a semi-amplitude, $C_3$ — is the moment of time corresponding to minimal value of the approximation, $C_4 = \frac{2\pi}{P} = \omega$ — angular frequency. If the signal is expressed in stellar magnitudes, the minimum of stellar magnitude $m$ corresponds to a maximal intensity $I$, according to the Pogson's law

$$m = m_0 - 2.5 \lg\left(\frac{I}{I_0}\right), \qquad (2)$$

where the index 0 corresponds to some standard star (e.g. [10]).

The test function, the minimum of which corresponds to the "best fit" parameters, may be written as [11]:

$$\Phi = \sum_{k=1}^{n} w_k p(u_k)(x_k - x_C(t_k))^2, \qquad (3)$$

where $w_k = \sigma_0^2/\sigma_k^2$ — is the weight of the observation corresponding to the accuracy of the measurement $\sigma_k$, $\sigma_0$ — is some positive constant, which is called the "unit weight error", $p(u_k)$ — a weight function, which is dependent on time difference $t_k - t_0$, but not on its accuracy $\sigma_k$. It is suitable to express $u_k$ in dimensionless units: $u_k = (t_k - t_0)/P$.

Here $t_0$ — is the "shift" in the wavelet terminology, i.e. the trial moment of time, for which the approximation is computed. The coefficients $C_1$ and $C_2$ may be computed using non-linear LS method, and $C_3$ and $C_4$ after some iterations using non-linear LS (differential corrections) and convergence to the values, which minimize the test function $\Phi$.

As the iterations for $C_3$ may converge not a minimum, but to a maximum, at each iteration, the value of $C_3$ should be corrected, if needed, by adding/subtracting $P/2$ to be inside the interval from $(t_0 - P/2)$ to $(t_0 + P/2)$.

The Eq. (1) may be rewritten in other forms using variables $u = (t - C_3)/P$ and $\varphi = 2\pi u$ for measuring time in units of the period and in radians, respectively:

$$x_C(t) = C_1 - C_2 \cos(2\pi \cdot (u - u_0)) = C_1 - C_2 \cos(\varphi - \varphi_0), \qquad (4)$$

The weight function used for the Morlet-type wavelet is a Gaussian

$$p(u) = \exp(-c \cdot u^2) = p_\varphi(\varphi) = \exp(-c_\varphi \cdot \varphi^2). \qquad (5)$$

Here $c$ is a non-negative constant, and $c_\varphi = c/4\pi^2$. For $c = 0$, $p(u) = 1$, and the approximation is not dependent on $t_0$, becoming a "global" one instead of a "local" one. The "classical" value is $c = 1/2$, so $c_\varphi = 1/8\pi^2 \approx 1/80$.

Small values $c \ll 1/2$ correspond to weighted asymptotically parabolic approximation

$$x_C(t) = (C_1 - C_2) + (2\pi^2 C_2/P^2) \cdot (t - C_3)^2 = \tilde{C}_1 + \tilde{C}_2 \cdot (t - C_3)^2. \qquad (6)$$

With $C_4 \to 0$, $P \to \infty$, $C_2 \to \infty$, $C_1 \to \infty$, as the coefficients of the parabola converge to "normal" values corresponding to $\tilde{C}_1 = x_C(C_3)$ and $\tilde{C}_2 = \ddot{x}_C(C_3)/2$.

In this case, no period may be determined.

A comparison of the results obtained using the wavelet analysis with different values of $c$ shows some "uncertainty principle" as the width of the peaks at the periodogram for a given shift $t_0$ is inversely proportional to an "effective duration" of the interval. So, the peaks are nearly constant in width at a logarithmic scale of periods, whereas the ordinary periodogram (asymptotically, for $c = 0$) has a nearly constant width for frequency $f = 1/P$ [3,4]. The "point-point" periodograms are reviewed in [12,13]. Examples of applications of the wavelet analysis to study semi-regular variables using various $c$ may be found e.g. in [14,15]. The catalogue of characteristics of 173 semi-regular variables was presented by [16].

The weight function (5) is an excellent choice for continuous data, which are infinite in length. However, there is no advantage for irregular discrete data because the LS approach makes possible to use many other functions. The exponent is a very time-consuming function and does not decrease the bias at the borders.

Andronov [11] proposed to use a compact weight function for generally aperiodic "Running Parabolae" (RP)

$$p(z) = (1 - z^2)_+^2 = \begin{cases}(1 - z^2)^2, & \text{if } |z| \leq 1, \\ 0, & \text{if } |z| > 1.\end{cases} \quad (7)$$

Here $z_k = (t_k - t_0)/\Delta t = u_k/\gamma$, and $\gamma = \Delta t/P$ is the parameter used e.g. in the "Running Sine" (RS) approximation (=wavelet with a fixed period and "rectangular shape", see [17] for a review).

Assuming the same value of the functions $p(z)$ and their first and second derivatives, $c = 1/(2\pi^2\gamma^2)$, and thus

$$p_e(z) = \exp(-2z^2). \quad (8)$$

Both functions are shown in Fig. 1. They are very close at small $|z|$, but have infinite and finite length of intervals of non-zero values, respectively.

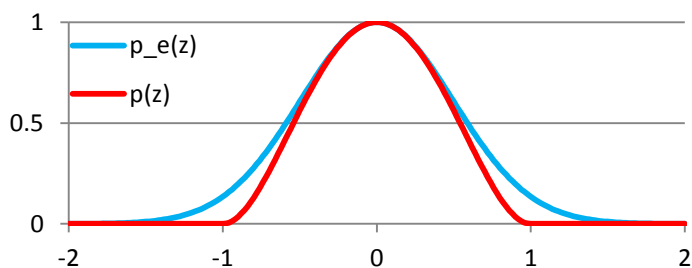

Fig.1. Weight functions $p_e(z)$ (Eq. (8)) and $p(z)$ (Eq. (7)).

Contrary to the RS algorithm with a "rectangular shape",

$$p_r(z) = \begin{cases}1, & \text{if } |z| \leq 1, \\ 0, & \text{if } |z| > 1,\end{cases} \quad (9)$$

the use of smooth functions (Eq.(8),(9)) tending to zero for large $|z|$ along with their derivatives causes corresponding smooth variations of the approximation and its derivative as well [11].

### Statistical properties of the approximation

To check statistical properties of the parameters, we generated $n_b = 10000$ "bootstrap" sets of $n = 101$ data points. The original "set" was defined as

$$x_{0k} = x_{00}(t_k) + \sigma_{00}\varepsilon_k \qquad (10)$$

$$x_{00}(t) = \exp(-\cos(\pi t)) \qquad (11)$$

where $t_{0k} = -2 + 4(k-1)/(n-1)$ are uniformly distributed arguments ranging from -2 to +2, the adopted "theoretical" period $P_{00} = 2$, $\varepsilon_k$ — are random numbers with theoretically "normal" (Gaussian) distribution with a zero mean and unit variance, and the "theoretical observational error" $\sigma_{00} = 0.2$. To make the signal not sinusoidal, an exponent was used in the theoretical function $x_{00}(t)$. The expected position of the minimum is $C_3 = 0$ and the scale coefficient $C_4 = \pi$. The initial data are shown in Fig. 2.

The basic functions for the differential corrections $f_\beta(t) = \partial x_C(t)/\partial C_\beta$ (see [3,4] for details), are

$$\begin{aligned} f_1(t) &= 1, \\ f_2(t) &= -\cos(C_4 \cdot (t - C_3)), \\ f_3(t) &= -C_2\, C_4\, \sin(C_4 \cdot (t - C_3)), \\ f_4(t) &= (t - C_3)\, C_2\, \sin(C_4 \cdot (t - C_3)). \end{aligned} \qquad (12)$$

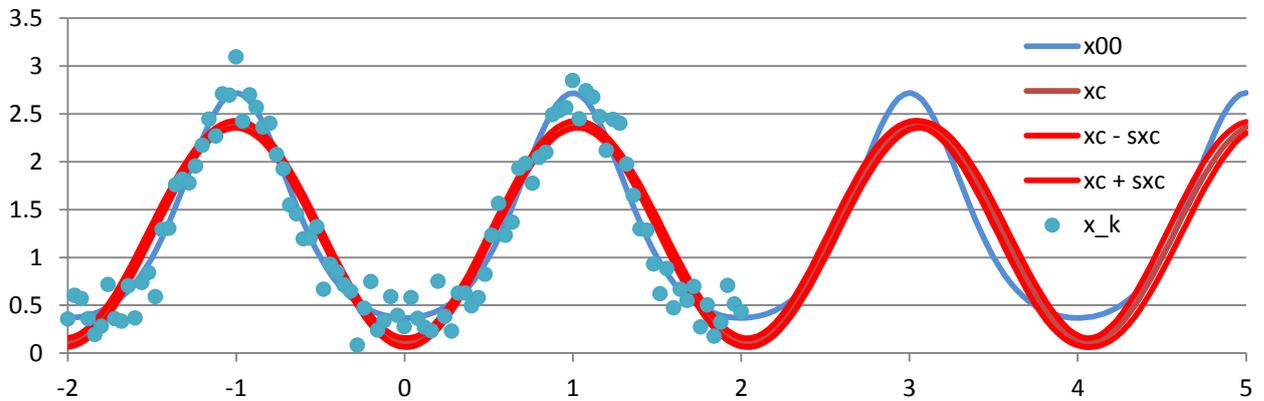

Fig. 2. The initial function $x_{00}$, the initial signal $x_k$, the approximation $x_C(t)$, and the "1σ error corridor" $x_C(t) \pm \sigma[x_C]$.

Occasionally, $C_2$ may be a "computer zero". In this case, the basic functions $f_3(t)$ and $f_4(t)$ are also zero. This makes degenerate the matrix of the normal equations. This problem may be solved by a shift of $C_3$ by $P/4$.

According to the bootstrap algorithm, from the initial set $t_{0k}, x_{0k}$, the artificial set $(t_k = t_{0j}, x_{0k} = x_j)$ is generated, where $j = \text{int}(n \cdot \text{rnd} + 1)$ is an integer randon number from 1 to n. Totally, $N_b = 10000$ random samples were generated, for which the parameters were determined using differential corrections.

Besides these parameters, we included additional parameters characterizing not the signal, but the random distribution of the numbers in the "bootstrap samples": $n_{min}, n_{max}$ – the minimal and maximal numbers of the observations; $n_{max} - n_{min}$ – the distance between the last and first observations in the sample; effective number of different observations [9,11]

$$n_{eff} = \frac{n^2}{\sum_{k=1}^{n} n_k^2}, \qquad (13)$$

$n_k$- is the number of occurences of the $k$−th data point from the original sample in the "bootstrap" sample; $n_0 -$ is the number of missing points; $n_m -$ is the maximal number of occurences of one point. Frequency polygons for these two characteristics are shown in Fig. 3. Unexpectedly, the dependencies are not smooth. As the sample mean values 51.0 and 36.9 and standard errors 3.5 are 3.2 for $n_{eff}$ and $n_0$, respectively. So the outliers may be due to sampling of discrete values of $n_0$ into the intervals. This suggestion is confirmed by comparison of the histogram of discrete values with the expected nearly Gaussian distribution.

We keep the Fig.3 with outliers, just to show this possible methodological problem, which may be solved by using discrete distribution instead of binning to subintervals.

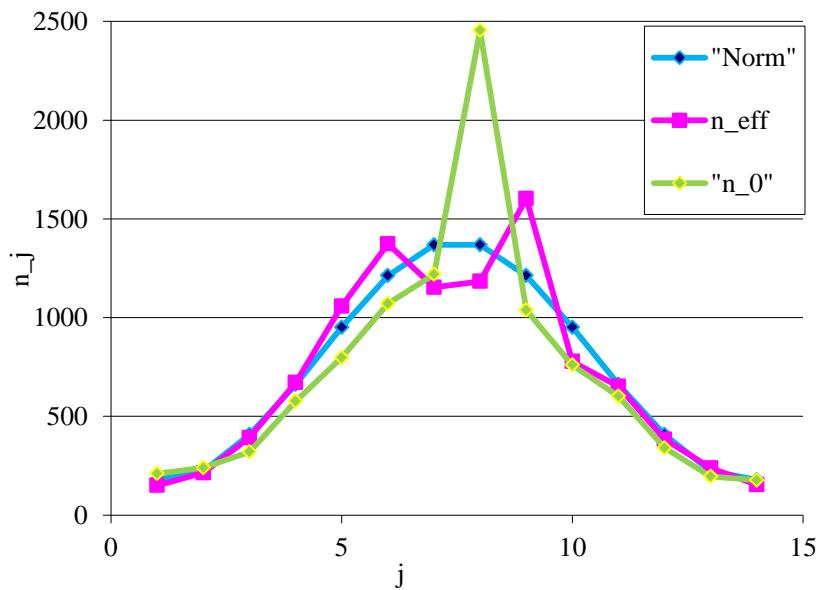

Fig. 3. Frequency polygons $n_j(j)$ as the number of points $n_j$ in a given $j$ − th interval for "Normal" (Gaussian) distribution, $n_{eff}$, $n_0$. The values of the parameters are centered to a sample mean and normalized by dividing by a standard error as $(x - \bar{x})/\sigma$.

The positions of the "random borders" $n_{min}$, $n_{max}$ may be shifted from the "natural borders" up to eight numbers (i.e. 0..8), with sample mean 1.56 and 100.43 (i.e. 0.56 from the borders of the original interval 1..101) and standard errors 0.92 and 0.95. The frequency polygons almost follow the exponential ("geometrical") distribution (Fig. 4):

$$n_j = (1 - e^{-1})e^{1-j}n. \tag{14}$$

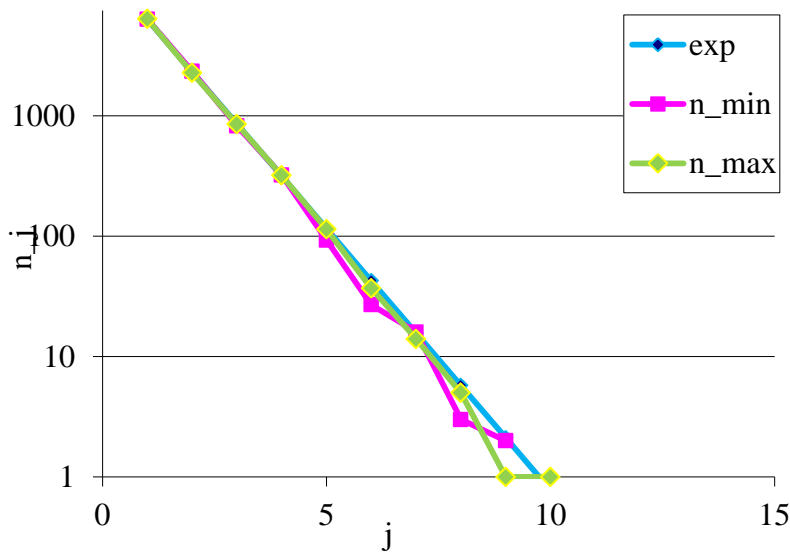

Fig. 4. Frequency polygons $n_j(j)$ as the number of points $n_j$ in a given $j-$th interval for a theoretical "exponential" distribution, $n_{min}$ and $n_{max}$.

The data-dependent parameters (after differential corrections) are: the coefficients $C_\beta$ and their error estimates $\sigma_\beta$; moment of extremum $t_e = C_3$, and corresponding extremal value $x_e = C_1 - C_2$; weighted r.m.s. deviation of the observations from the approximation $\sigma_W$; unit weight error $\sigma_0$.

In Fig. 4 are shown frequency polygons only for two parameters – with a good agreement with normal distribution for the moment of extremum $t_e$, and some systematical deviation and corresponding extremal value $x_e$. This may be explained by an asinusoidal shape of the function $x_0(t)$, which has maxima, which are sharper than the minima. The cosine approximation is systematically lower at the minimum than the "pure signal" $x_0(t)$ or a "noisy data" $x_k$. So the shift and asymmetry of the distribution are caused by systematic difference in shape between the signal and approximation. This effect is much smaller for the parameters $C_\beta$.

The error estimates of the accuracy of parameters obtained using the "bootstrap" samples are significantly larger than that obtained using the least squares method, by a factor ranging from 1.31 to 1.41. A simple estimate of the ratio $((n-m)/(n_{eff}-m))^{1/2} = 1.44$ is in a reasonable agreement with the numbers determined above.

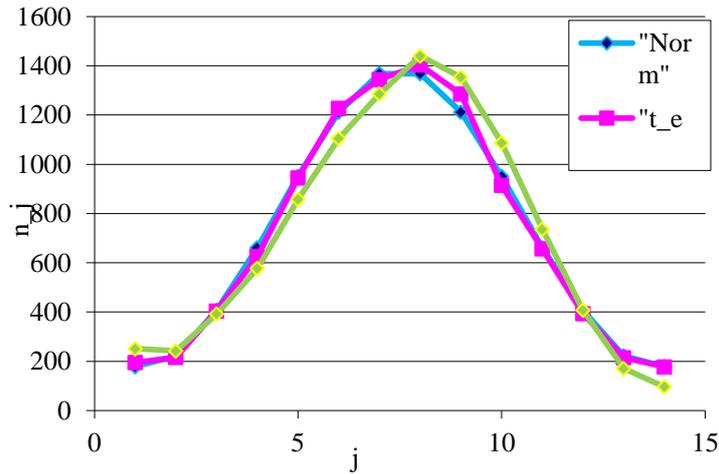

Fig. 4. Frequency polygons $n_j(j)$ as the number of points $n_j$ in a given $j$ — th interval for the normal distribution ("Norm"), moment of extremum $t_e = C_3$, and corresponding extremal value $x_e = C_1 - C_2$. The frequency polygons for "Norm" and $t_e$, practically coincide with each other, but not with that of $x_e$.

There are shifts between the parameters obtained for the original data sample (all points occur once) and the mean values for the same parameters obtained using "bootstrap" data. Except $n_{eff}$, $n_0$, for other parameters, the parameter $(x - \bar{x})/\sigma$ (parameter from the original sample minus the mean for bootstrap estimates, divided by a standard error from the bootstrap parameters) is typically much less than unity. However, the standard error of the mean is by a factor of $\sqrt{n}$ less than the standard error of the data. In this case, the difference becomes much more significant.

Next question is on estimate of the statistical error of the moment minimum (and possibly other parameters) using the "bootstrap" technique (initially introduced by [18,19]. Brát et al. [20] described their program based on the function separately studied by Mikulášek [21]. The output contains two values $\sigma_- = t_e - t_{0.025}$ and $\sigma_+ = t_{0.975} - t_e$, , which correspond to percentiles $t_{0.025}$ and $t_{0.975}$, i.e. removing 2.5% of estimates either from small, or large side of the distribution.

This challenges the typical single definition of $\sigma$ as an error estimate, which should be the same for positive and negative directions [22]. Moreover, this is used for the weight in Eq. (3). Following [13], one may argue that, assuming the Gaussian continuous distribution of observational errors and their statistical independence [3], $\sigma_- = \sigma_+ = 1.96\sigma$ for this probability level 95%. This means a decrease of the statistical weight $w_k$ by a factor of $1.96^2 \approx 3.84 \approx 4$ рази.

Thus one has to define a function $\sigma(\sigma_-, \sigma_+)$, which converts two values $(\sigma_-, \sigma_+)$ to a single $\sigma$. There may be recommended similar methods of averaging: the r.m.s. $\sigma \approx ((\sigma_-^2 + \sigma_+^2)/2)^{1/2}/1.96$; the value corresponding to a mean weight $\sigma \approx ((\sigma_-^{-2} + \sigma_+^{-2})/2)^{-1/2}/1.96$, or to an interpercentile interval $\sigma \approx (\sigma_- + \sigma_+)/2/1.96$. If $\sigma_-/\sigma_+$ is close to unity, all three approximations are very close to each other.

This is also seen in our numerical experiment described above. Much more important to take into account the scaling by 1.96, which is not taken into account in the popular software described by [20].

If $\sigma_-/\sigma_+$ is far from unity, the distribution is non-Gaussian, and thus the use of the weights may be done, but loses its statistical justification as that corresponding to maximum likelihood [22].

Similarly, one may use other percentiles, e.g. that corresponding to $1\sigma$:

$$\sigma \approx (t_{0.841} - t_{0.159})/2. \tag{15}$$

Here the percentiles correspond to a probability $1 - 2 \cdot 0.159 = 0.682$ instead of arbitrary 95% mentioned above. From our numerical experiments, both interpercentile estimates coincide with "bootstrap" r.m.s. value of $\sigma$ within few per-cent (except variables with an exponential distribution). So they both may be recommended. However, the software [20] produces only $\sigma_-$ and $\sigma_+$, so

$$\sigma \approx (t_{0.975} - t_{0.025})/2/1.96 = (\sigma_- + \sigma_+)/2/1.96 \tag{16}$$

As was mentioned above, this value is anyway larger than the LS estimate by a factor of 1.3…1.5, which causes an apparent additional decrease of the weight by a factor of $\approx 1.7 \ldots 2.2$.

**Discussion**

We have tested numerically the modification of the wavelet analysis using a compact (time-limited) weight function using non-linear least squares and an alternate "bootstrap" method of estimating statistical errors of the parameters.

Obviously, the bootstrap may be used for other types of approximations, for which we used the LS estimates: e.g. "global" trigonometric polynomials of statistically optimal order [3,23,24], "local" algebraic polynomials of statistically optimal order [16,25], "symmetrical polynomials" [26], polynomial splines [27], "wall-supported" functions [28], "New Algol Variable" [29-33], "asymmetric hyperbolic secant" [34] and some other methods [35-36].

The software MCV [37] is oriented mainly on "global" approximations, whereas MAVKA [38,39[ is oriented to statistically optimal determination of moments of extrema (ToM="Time of Minimum"), in the AAVSO [40] terminology. In MAVKA, the total number of approximations is 11 (totally, 21 function), but the wavelet analysis is not included yet in these programs.

The wavelet analysis is an effective tool for semi-regular pulsating variables [41-43], symbiotic binaries [44] and cataclysmic variables [36].

The proposed local weight function makes the wavelet analysis faster.

**Acknowledgements.**

This work was initiated by Dr. Bogdan Wszołek, who makes a great contribution to public outreach and popularization of astronomy via the international society "Astronomia Nova".

This work is a of the "Inter-Longitude Astronomy" [45,46] international project, as well as of the `'Ukrainian Virtual Observatory" [47,48] and "AstroInformatics" [49].